\documentclass[final,5p,times,twocolumn,longbibliography]{elsarticle}

\usepackage{graphicx}
\usepackage{dcolumn}
\usepackage{bm}
\usepackage{hyperref}
\usepackage{amsmath}
\usepackage{xcolor}

\journal{Physics Letters B}

\begin{document}

\begin{frontmatter}

\title{\boldmath Experimental Impact of Jet Fragmentation Reference Frames At Particle Colliders}

\author[first]{Lawrence Lee} 
\author[first]{Charles Bell} 
\author[first]{John Lawless} 
\author[first]{Cordney Nash} %
\author[first]{Emery Nibigira} 

\affiliation[first]{organization={University of Tennessee, Knoxville},
            addressline={1408 Circle Dr}, 
            city={Knoxville},
            postcode={37996}, 
            state={TN},
            country={USA}}


\date{\today}

\begin{abstract}
In collider physics, the properties of hadronic jets are often measured as a function of their lab-frame momenta. However, jet fragmentation must occur in a particular rest frame defined by all color-connected particles. Since this frame need not be the lab frame, the fragmentation of a jet depends on the properties of its sibling objects.  This non-factorizability of jets has consequences for experimental jet techniques such as jet tagging, boosted boson measurements, and searches for physics Beyond the Standard Model. In this paper, we will describe the effect and show its impact as predicted by simulation.
\end{abstract}


\end{frontmatter}

\section{Introduction}

At particle colliders, collimated sprays of hadrons known as jets are commonly produced. Roughly, jet activity is the collider signature for parton production in scattering processes. QCD confinement forbids free particles from carrying color charge such that a fragmentation and hadronization process yields an observable jet of color-singlet hadrons. This process results in small angle particle production, which leads to the observed collimation of fragmentation products.

Experiments measure the momenta and properties of these hadrons and cluster the reconstructed objects into jets using various algorithms~ \cite{Dokshitzer:1997in, Cacciari:2008gp}. The properties of the aggregate jet objects are roughly used as a proxy for the properties of the initiating partons in some leading-order approximation~\cite{Komiske_2018vkc}. In this approximation, an initiating parton will fragment into additional partons with energies as dictated by non-perturbative fragmentation functions. The overall particle multiplicity and the distribution of energy across these particles is stochastically determined by these fragmentation functions. 

Beyond the measurement of a four-momentum, modern collider experiments frequently measure the internal structure of a jet, usually to determine the ``origin'' of the jet. Classical methods to determine if a jet originated from heavy-flavor quarks or $\tau$ leptons long predate the Large Hadron Collider (LHC)~\cite{Bandurin:2014bhr}. Within the LHC era, additional tools such as quark vs. gluon ($q/g$) tagging and the industry of jet substructure continue to use the distribution of jet constituents as a discriminating tool \cite{Larkoski_2019,Frye_2017,Komiske_2018vkc, Kogler:2018hem}. Both angular and longitudinal distributions of energy flow can carry crucial information about the origin of a jet~\footnote{More unconventional signatures can also lead to anomalous distribution of energy within a jet due to displaced decays, showering via a hidden dark sector, and via other mechanisms~\cite{Lee:2018pag,Alimena:2019zri,Cohen:2017pzm}.}.

In the jet physics industry, it is common to consider a jet to be defined by its progenitor particle species and its momentum; \textit{e.g.}, a light quark jet of a particular momentum should have a set of properties drawn from the same distributions as every other light quark jet of that momentum scale. Each jet is assumed to have a largely independent fragmentation process, an assumption we will call \textit{jet individualism}, modulo small-but-measurable colorflow effects~\cite{ATLAS:2018olo,ATLAS:2022ctr,CMS:2022awf}. We often obtain independent samples of jets in a particular momentum range from other physical processes in data to develop data-driven background estimates, building flavor- and momentum-dependent templates for jet structure distributions~\cite{Cohen:2014epa, ATLAS:2015xmt,ATLAS:2014vax,CMS:2021iwu}. In flavor-tagging, it is often stated that an experiment has a particular tagging performance for $b$-jets with a given momentum.

In this paper, we will present an argument that, in retrospect, seems obvious: The ubiquitous experimental assumption that jets of a given flavor and momentum can be thought of as independent, identically-distributed objects is invalidated by a simple, first-principles consideration. We describe a hadronic jet's color rest frame dependence, example processes where this effect can be observed, and how different event generators model these effects. We then identify techniques where the effect is under-appreciated at modern experiments and explain a potential strategy to use this effect for process discrimination.

\begin{figure*}[tbp]
    \centering
    \includegraphics[width=0.7\textwidth]{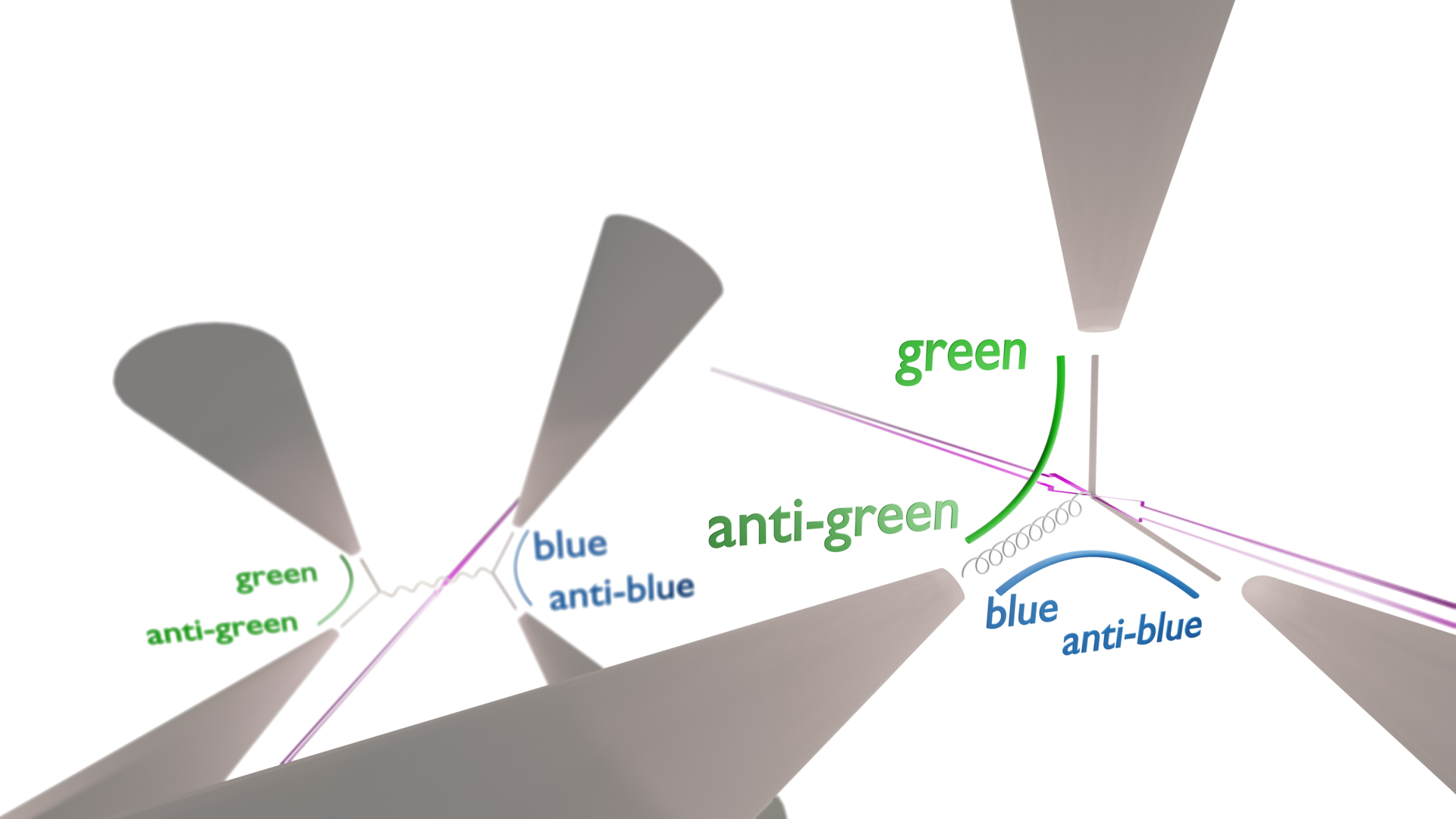}
    \caption{Illustration of some example color-flow topologies at a lepton collider. (Left) The collision of two leptons (purple) is producing two color singlets, each decaying to a $q\bar{q}$ pair. The color rest frames are the frames where each of the color-connected $q\bar{q}$ systems is at rest. (Right) Another collision of leptons producing a ``Mercedes star'' trijet event where the $q\bar{q}$ system is recoiling off of a high momentum gluon. In this case, all three initial partons are color connected, and the color frame is the same as the lab frame. Cones schematically represent the fragmentation and hadronization process that gives a jet.}
    \label{fig:colorflowCartoon}
\end{figure*}

\section{Jet Individualism}

Jet properties are often discussed as a function of the jet's momentum or energy as seen in the detector. We think about the fragmentation properties of an $X$~GeV jet as a function of its kinematics. In collider data analysis, it is common to find ``standard candles'' of kinematically similar jets from well-understood processes for tagger construction. For example, we can inform jet measurements using well-balanced dijet events and use that understanding in searches for a new particle decaying to jets with similar kinematics~\cite{ATLAS:2023tyv,CMS:2016lmd}.

This way of thinking about jets relies on the assumption that kinematically similar jets of similar origin are all produced from the same underlying physics distributions. The angular distribution of particles, the energy sharing between them, and the total number of final state particles in a jet are assumed to be entirely dictated by the kinematics and origin of the jet. However, this jet individualism assumption and our ordinary language in jet physics use the lab frame momentum to discuss these properties.

This assumption fails once requiring that observers in all frames must have a consistent view of these jets. Since one can boost into a frame where that lab-frame $X$~GeV jet is now $Y$~GeV, not all the properties of a jet can be solely defined by its lab-frame momentum. In principle, these properties are defined by all color-connected objects and do not need to directly depend on lab frame momenta.

The clearest example of this effect is in a popular jet observable -- the particle multiplicity~\cite{ATLAS:2014vax, CMS:2021iwu, Frye_2017, ATLAS:2019rqw, ATLAS:2016vxz}. While measuring this by counting the number of charged particle tracks within the jet $n_{trk}$ is infrared and collinear (IRC) unsafe, it's a common observable for jet discrimination in practice. It's well established that the average $n_{trk}$ of these jets is strongly dependent on momentum scale; jets at larger momentum contain more particles~\cite{ATLAS:2019rqw}. However, since observers in all Lorentz frames must agree on the \emph{number} of particles produced in the fragmentation process, there must be a preferred frame in which the fragmentation function is valid. To understand the particle multiplicity of a jet, not only is the jet's momentum needed, but all jets color connected to its ``initiating'' partons need to be considered when deciding the proper rest frame for fragmentation.

Different shower models use different choices of reference frames to fragment a jet. 
\begin{itemize}
    \item In the Pythia event generator~\cite{Sjostrand:2014zea,Bierlich:2022pfr}, the basic showering process uses several models which work in several frames, often the color dipole rest frame. Pythia's hadronization model, the Lund String model, hadronizes the strings in the rest frame of all color-connected strings.
    \item Vincia~\cite{Fischer:2016vfv} is a newer shower model for Pythia, that is now the default. In Vincia, the main model used is their antenna model, which is similar to the dipole picture. The reference frame handling is similar to that of the previous Pythia model.
    \item In the Herwig model~\cite{Bahr:2008pv,Bellm:2015jjp}, a coherent branching algorithm is used to perform parton showering. The parton shower occurs ``in the rest frame of the progenitor and an object with which it shares a color line''~\cite{Bahr:2008pv}. To hadronize partons, Herwig then uses the cluster model. In this model, after gluon splitting and cluster fission, hadronization occurs in a cluster's own rest frame.
\end{itemize} 

In this paper, we will consider the \emph{``color rest frame''} (or maybe ``\emph{center of color}'' frame). Instead of considering individual jets in the lab frame, we consider jet fragmentation in the rest frame of all color-sibling particles. The specifics of this simple assumption do not affect our conclusions.

This Frame of Fragmentation and Showering (FFS) effect is at odds with many of the assumptions in modern experimental jet physics at the LHC where lab frame fragmentation is assumed and process- and color-flow-dependent effects are usually ignored~\footnote{The effects of color flow on the jet pull observable have been measured at the LHC~\cite{ATLAS:2018olo,ATLAS:2022ctr,CMS:2022awf}. In these measurements, the fragmentation patterns of pairs of jets is used to infer parton-level color connection.}. This effect can lead to large differences in fragmentation patterns in very common final states. A simple illustration can be seen in the collision of two color-singlet fundamental particles, like those at lepton colliders. In simple multijet final states at symmetric $e^+e^-$ colliders, the color rest frame is the same as the lab frame, since all outgoing partons are color connected and there is no net momentum in the lab frame. However, if two jets are produced in the decay of a $Z$ boson, the color rest frame is the rest frame of the $Z$, which may be significantly boosted with respect to the lab frame. The fragmentation of these jets is set by the energy scale $m_Z/2$ and has no dependence on the energy measured by the detector or the boost of the $Z$~\footnote{This was explicitly noted in Ref.~\cite{ATLAS:2015irp}.}. The same is true for any hadronically-decaying color singlet.

In analyses looking for boosted vector bosons $V$ decaying hadronically,  the signal jet fragmentation is set by $m_V/2$, whereas multijet backgrounds most likely have complex color connections with the whole final state (and the beam remnant at hadron colliders), which leads to very different underlying fragmentation distributions. In the case of the ``Mercedes" star topology that led to the discovery of the gluon, two quarks and one gluon are produced. The color octet gluon is color connected to both of the color triplet quarks. In such cases, the color rest frame is the rest frame of all three jet objects, which at a lepton collider, will be the same as the lab frame. The difference between these color topologies is illustrated in Figure~\ref{fig:colorflowCartoon}. The effect will be more difficult to understand at hadron colliders like the LHC. In QCD multijet processes at hadron colliders, in addition to potential color connections between  final state jets, color connection to high-energy beam remnants can lead to very highly boosted color rest frames.

To investigate how the color rest frame dependence of jet fragmentation is handled by Monte Carlo generators, samples were produced with various parton shower models. MadGraph5\_aMC@NLO~3.3.1~\cite{Alwall:2014hca} was used to produce 50,000 parton-level events of two processes: $e^+e^- \to 3j$ and $e^+e^- \to ZZ \to 4j$, both with a collision energy of $\sqrt{s}=1$~TeV. In the first process, the color rest frame is coincident with the lab frame, such that jet observables show an expected lab-frame momentum dependence. In the second, the color rest frames are always the rest frames of the $Z$ bosons such that some jet observables will be set by $m_Z/2$ and not scale with the lab-frame momentum.

These parton-level events were then showered using three different models: Pythia 8.306~\cite{Sjostrand:2014zea,Bierlich:2022pfr} using ``simple showers,'' Vincia~\cite{Fischer:2016vfv}, and Herwig~7.2.2~\cite{Bahr:2008pv,Bellm:2015jjp}. Parton shower matching is not included. Since these effects are on jet-level observables, the rate of extra jet production should not affect our conclusions, and potential double counting of hard radiation effects do not affect the color handling in the generators. Jet clustering was performed at particle level using Fast-Jet~\cite{Cacciari:2011ma} as interfaced in the Delphes 3.5.1~\cite{deFavereau:2013fsa} framework. These particle-level truth jets were clustered using the anti-kt algorithm with an $R=0.4$ radius parameter~\cite{Cacciari:2008gp}. To decouple the FFS effect from the fragmentation differences between quarks and gluons, gluon-initiated jets are not considered in these studies by requiring that the highest momentum matched parton is not a gluon. For similar reasons, $b$-jets are also excluded.

\begin{figure}[tbp]
    \centering
    \includegraphics[width=0.45\textwidth]{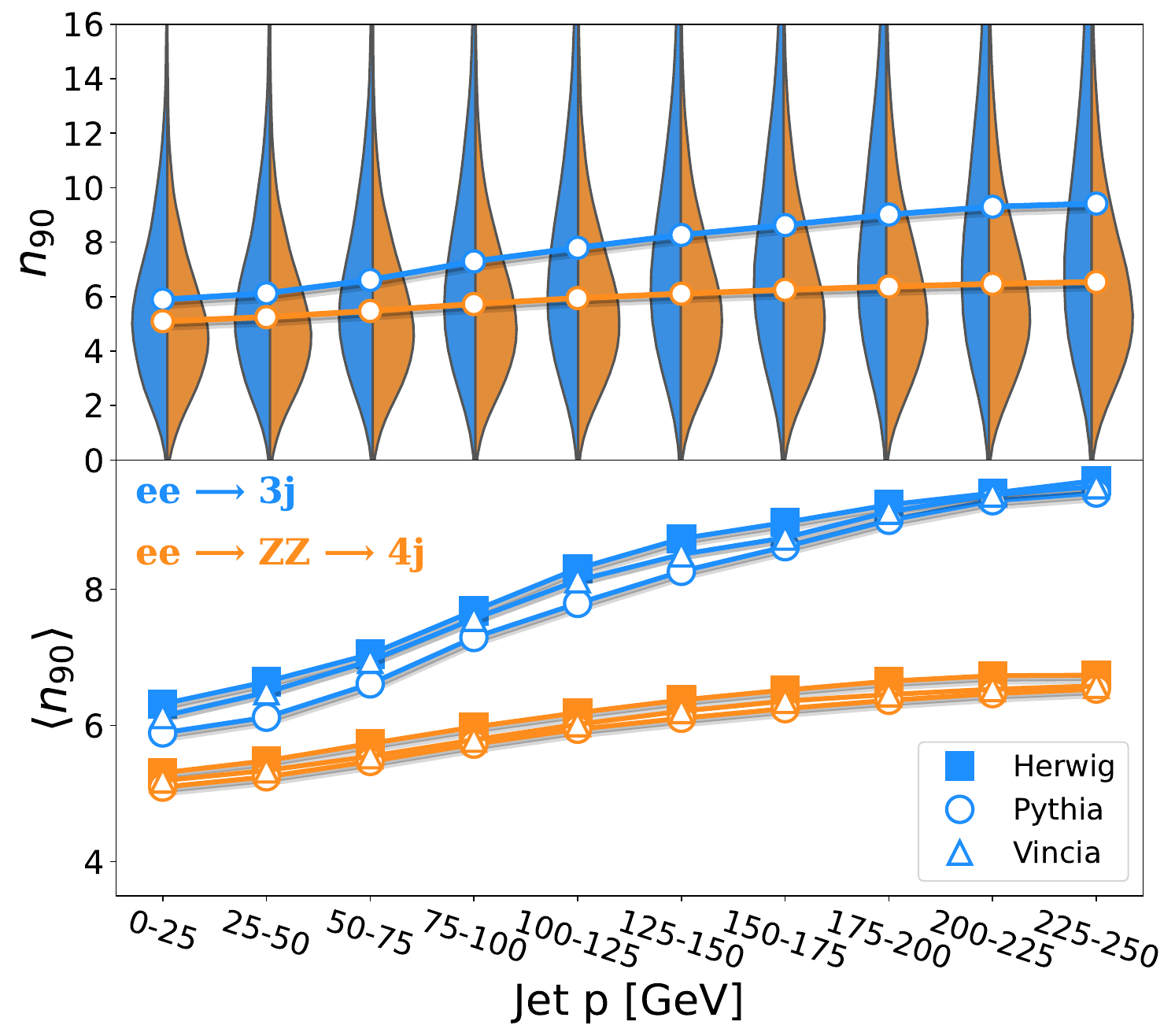}
    \caption{(Top) Split violin plots showing the normalized $n_{90}$ distributions for quark-initiated jets, as modeled in Pythia using the ``simple shower'' model, for two $ee$ collider processes as a function of lab-frame momentum. In $ee \to 3j$ (blue, left), the color rest frame and lab frame are coincident, and the particle multiplicity is a function of the momentum. However, in $ee \to ZZ \to 4j$ (orange, right), in which jets obtain large momentum from boosted color rest frames, the dependence is significantly weaker. The means of each distribution are shown as markers. (Bottom) The mean $\langle n_{90}\rangle$ is shown as a function of lab-frame jet momentum for multiple shower models. Herwig, Pythia, and Vincia show similar behaviors. All samples use the same matrix element events such that they are statistically identical up to the parton shower model. 
    }
    \label{fig:n90}
\end{figure}

We define $n_{x}$ as the minimum number of particles in a jet that give $x\%$ of the jet's total momentum. A per-jet interpolation is performed to define a fractional multiplicity. The jet constituents are ordered by decreasing momentum and cumulatively summed. The fractional number of particles needed to recover $x\%$ of the total jet momentum is used. Figure~\ref{fig:n90} shows $n_{90}$, the minimum number of particles that could account for 90\% of the jet's energy as a function of the lab-frame momentum of the jet. In this figure, only quark-initiated jets are considered in order to remove any effects from a momentum- or process-dependent gluon fraction. While $n_{90}$ varies with the lab-frame momentum for $e^+e^- \to 3j$, this dependence is significantly weaker in the $e^+e^- \to ZZ \to 4j$ case. Figure~\ref{fig:n90} also shows the lab-frame momentum dependence of the average multiplicity $\langle n_{90}\rangle$, comparing the three shower models considered. Since each sample uses the same matrix element level events, statistical uncertainties can be ignored up to the parton shower level. All considered models demonstrate the FFS effect in the process dependence of the mean at high momentum. While small differences exist, the qualitative behaviors are similar. For jets with a measured momentum of 200 GeV, differences in $\langle n_{90}\rangle$ as large as roughly 50\% are predicted for different color topologies; this FFS effect can be sizable at typical collider energy scales.

\begin{figure}[tbp]
    \centering
    \includegraphics[width=0.45\textwidth]{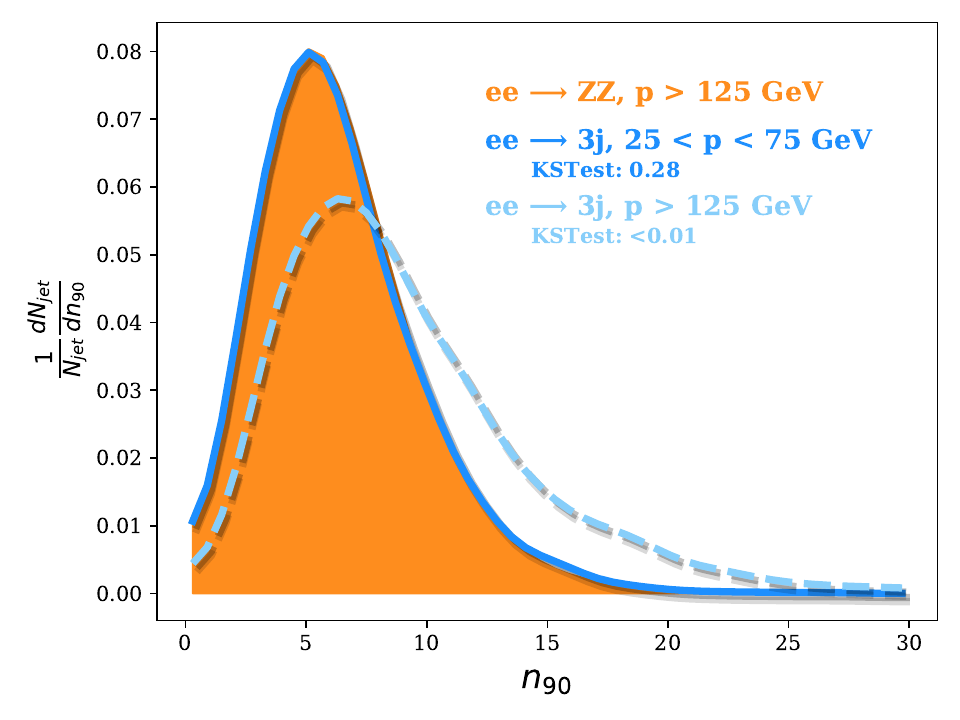}
    \caption{Jets in a momentum region above 125 GeV produced in $e^+e^- \to ZZ \to 4j$ are shown in orange to show the asymptotic behavior of this sample in Figure~\ref{fig:n90}. The two blue curves show the behavior of the $e^+e^- \to 3j$ sample at two different values of the lab-frame momentum. The high momentum orange distribution is highly compatible with the low momentum $e^+e^- \to 3j$ sample roughly centered around $m_Z/2$ (solid blue) with a KS test p-value of 0.28. In contrast, this sample in the same $>125$~GeV region yields a different distribution (dashed light blue) with a KS test p-value consistent with 0.
    }
    \label{fig:fixed-zz}
\end{figure}

In Figure~\ref{fig:n90}, jets from the $e^+e^- \to 3j$ process show the conventional increase of $n_{90}$ as a function of momentum. In contrast, the $e^+e^- \to ZZ \to 4j$ distribution is at significantly lower $n_{90}$ values and shows some initial growth before the mean asymptotically approaches a constant value at high momentum. This asymptotic limit is consistent with the $e^+e^- \to 3j$ jets at a momentum scale of around $45$ GeV, as expected when accounting for the FFS effect. Figure~\ref{fig:fixed-zz} shows the high momentum jets from the $e^+e^- \to ZZ \to 4j$ process compared to  two different momentum slices from the $e^+e^- \to 3j$ case. We observe consistency between the asymptotic behavior of the $e^+e^- \to ZZ \to 4j$ case and the $25\mbox{--}75$~GeV range of the $e^+e^- \to 3j$ case, with a KS test $p$-value of $0.28$. 
These jets from the $e^+e^- \to ZZ \to 4j$ process are showering as if they have a momentum of $m_Z/2$. The distribution of jets from the  $e^+e^- \to 3j$ process at comparably high lab-frame momenta differ significantly, with a KS test $p$-value of $<0.01$.

The low momentum behavior of the $e^+e^- \to ZZ \to 4j$ distribution in Figure~\ref{fig:n90} can be explained by additional complexity in the decay of the Z boson. For example, low momentum jets in this sample have a sizable contribution from additional hard gluon emissions, producing additional, softer jets.

\section{Implications for Modern Experimental Collider Data Analysis}

Despite the fact that this effect is modeled in modern simulations, the typical practicing collider experimentalist ignores it when designing experiments and analyses. The assumption that a light jet of 100 GeV is a light jet of 100 GeV will break down from a color-connection dependence effect. This has implications for many of the jet techniques used in modern collider physics.

In particular, this FFS effect has significant implications on how jet tagging is done. Jet tagging algorithms try to gain insight into the origin of a jet using the observable properties of its shower. In the design, training, and validation of these taggers, the jet individualism assumption is heavily used.

The tagging method with the largest tradition in hadron collider physics is $b$-tagging. Simple $b$-taggers use the large lifetime of B hadrons to identify displaced decay products within a jet, by observing large impact parameter tracks or displaced track vertices. Modern $b$-taggers use much more information about the jet, often employing deep neural networks with a large amount of low-level information about the jet, such that these taggers are likely sensitive to FFS effects~\cite{ATLAS:2019bwq,CMS:2017wtu}.

Training $b$-taggers at high $p_T$ without controlling for the FFS effect can lead to incorrect interpretation of selection efficiencies or non-optimal performance. In Ref.~\cite{ATLAS:2019bwq}, the ATLAS experiment describes their $b$-tagging training procedure: ``... the new Run 2 $b$-tagging algorithm training strategy is based on the use of a hybrid sample composed of both the baseline $t\bar{t}$ event sample and a dedicated sample of $Z'$ decaying into hadronic jet pairs.'' This $Z'$, referring to a heavy, color singlet, $Z$-like resonance, decays to $b\bar{b}$ pairs in these samples. The $Z'$ mass is set to $4$~TeV and a large range of jet momenta are populated. For nonzero $Z'$ momentum, the lab frame momentum of the jets and the color rest frame momentum will differ. The fragmentation will be pinned to the $m_{Z'}/2$ scale and will not represent $b$-jets from other processes at similar lab-frame momentum scales. Especially as modern taggers use more information about the jets in advanced machine learning techniques, these tools will become more sensitive to the FFS effect in ways that are difficult to control.

Distinguishing quark-initiated and gluon-initiated jets is a popular technique at modern colliders~\cite{ATLAS:2014vax,CMS:2021iwu,Gallicchio:2011xq,Larkoski_2019}. Early in the LHC, these $q/g$ tagging techniques used the track multiplicity, jet width, and other characteristics to differentiate these jets, taking advantage of the gluon's larger color factor. Modern techniques include using lower level information with machine learning techniques. LHC experiments build templates of these variables in dedicated quark- and gluon-enriched control regions to inform taggers used in data analyses, particularly those looking for quark-dominated processes~\cite{Gallicchio:2011xc}. Despite some structural differences between quark and gluon jets, the differences are subtle and vary greatly with momentum~\cite{ATLAS:2014vax,CMS:2021iwu}. Since the fragmentation will be impacted by the FFS effect, jets from the decay of a boosted color singlet can be easily misidentified, particularly in decays involving gluons.

These $q/g$ discrimination efforts have assumed jet individualism, without regard to sibling jets. Community-wide $q/g$ discrimination challenges have published shared ntuple datasets at the jet level and not the event level~\cite{IMLQGChallenge}. A host of ML techniques have been engineered to take the properties of individual jets as inputs. A recent experimental review of jet substructure at the LHC, Ref.~\cite{Kogler:2018hem}, when discussing the jet inputs used for a $q/g$ discriminator, uses the phrase ``Since the distributions of these variables depend on $\eta$, $p_T$,...'', which emphasizes the lab-frame, jet individualistic assumptions throughout the field~\cite{Metodiev:2017vrx}.

As put clearly in Ref.~\cite{Metodiev:2018ftz}, ``... there are well-known caveats to this picture of jet generation, which go under the name of `sample dependence'.'' Such event-level effects are then argued to be small \cite{Metodiev:2018ftz, ATLAS:2014vax,Bright-Thonney:2018mxq,Komiske_2018vkc} and subsequently ignored. ``Here, we assume that sample-dependent effects can either be quantified or mitigated ...'' \cite{Metodiev:2018ftz}. We argue that there exist sizable sample-dependent effects that are neither simple to quantify nor to mitigate.

\begin{figure}[tbp]
    \centering
    \includegraphics[width=0.45\textwidth]{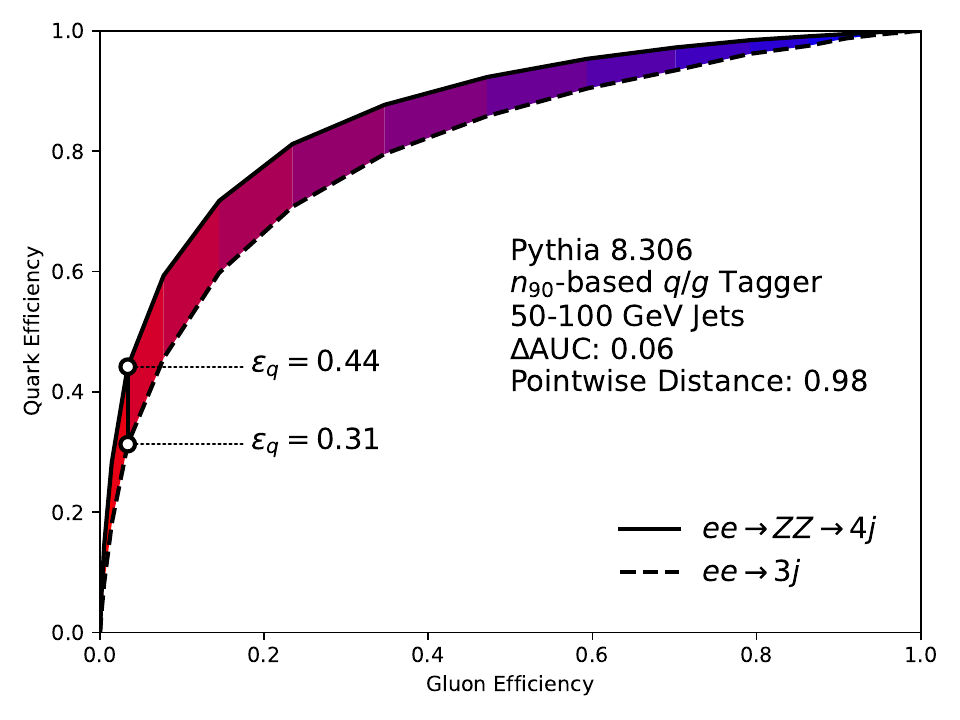}
    \caption{
        A ROC plot showing the tagging efficiency for a $q/g$ tagger based on a maximum $n_{90}$ requirement. In the dashed line, the performance is shown for differentiating quark and gluon jets in the $ee\to3j$ process. In contrast, the performance is also shown (solid line) for differentiating quark jets that originate from a $ee\to ZZ\to 4j$ process from the gluon jets that originate from $ee\to 3j$ background processes. While the areas under the ROC curves (AUCs) differ by only 6\%, the actual performance of a given working point will vary much more across these samples. Colors in the band represent potential working points of the tagger.
    }
    \label{fig:qgroc}
\end{figure}

To demonstrate the possible effect of this sample dependence on an analysis, we designed a simple $q/g$ tagger, identifying quark jets as those with an $n_{90}$ value below a threshold. We constructed the tagger using $n_{90}$ templates derived from trijet events, but then use the tagger on the $ee\to ZZ\to 4j$ events discussed above. This is a real-world example that represents a common use for modern $q/g$ tagging: trying to find the quark jets from the decay of a color singlet while rejecting gluon jets from background processes that do not include intermediate color singlets. For a given $n_{90}$ requirement, the tagger is expected to have an efficiency for identifying quark jets $\epsilon_q$ and an efficiency for misidentifying gluon jets as quark jets $\epsilon_g$, both obtained by the fraction of the $n_{90}$ distribution below the threshold. Scanning the value of the $n_{90}$ threshold produces the ROC curves shown in Figure~\ref{fig:qgroc}. These efficiencies are calculated using the Pythia8 parton shower model.

If the tagger templates are constructed using jets collected in $ee\to 3j$ events, we find a 31\% quark tagging efficiency and a 3.4\% gluon tagging efficiency (\emph{i.e.} successfully rejecting 96.6\% of gluon jets), as highlighted in Figure~\ref{fig:qgroc}. Assuming jet individualism, we might then take this tagger and assume this performance should apply in an analysis looking for the $ee\to ZZ\to 4j$ process. However, because of the FFS effect, the quark $n_{90}$ distribution in the $ee\to ZZ\to 4j$ process is at significantly lower values for the same lab-frame momentum, as demonstrated in Figure~\ref{fig:fixed-zz}. Instead of a 31\% quark jet efficiency, this tagger gives a 44\% quark jet efficiency in such events for the same $n_{90}$ threshold. The resulting quark jet sample is 40\% larger than expected from the initial tagger construction. This sample dependence is commonly ignored and is well outside of the systematic uncertainties typically used in contemporary experiments.

The overall performance quantified by the full ROC curve is minorly affected by the FFS effect, with a difference of only 0.06 in the area under the curve (AUC) metric. However, this measure characterizes the tagger in its potential performance after optimization of a working point. The actual performance impact of a particular working point, for example the markers shown in Figure~\ref{fig:qgroc}, is much larger than the AUC difference suggests. Individual ROC points do not map onto nearby points, and there is instead, a nonzero shear in the mapping of one ROC curve to the other.

A particularly extreme example of how the FFS effect affects the goals of $q/g$ tagging efforts is in the tagging of jets from vector boson fusion (VBF) and scattering (VBS) processes. At hadron colliders, VBF/VBS processes give rise to two quark jets that are, at leading order, color connected to only the beam remnants, and to nothing else in the event. Backgrounds for analyses looking to measure these processes often give pure QCD (\textit{i.e.}, gluon enriched) jets such that $q/g$ discrimination methods are attractive and often used~\cite{ATLAS:2020bhl,ATLAS:2018jvf,CMS:2015ebl}.

However, each quark jet that emerges from the VBF/VBS process will fragment in the color rest frame of the highly boosted color dipole formed with the relevant beam remnant. At the LHC, these frames are highly boosted with respect to the lab frame, and in detailed fragmentation observables, these jets will look nothing like the typical jets used to train and characterize such taggers. To successfully $q/g$-tag VBF/VBS processes, dedicated template distributions would need to be derived or existing templates would need to be altered to account for the FFS effect and somehow control for the beam remnant momentum.

These effects can have an impact any time two different color topologies are being compared. Using only properly Lorentz-transforming observables is one method that is certain to prevent the FFS effect from being an issue. Alternatively, adversarial training of discriminants could help ensure that observables are insensitive to this effect. In general, jets cannot be simply taken out of the context of their event. The sample dependence must be considered to combat the FFS effect. 

On the other hand, the FFS effect could be exploited for process discrimination. In searches for boosted color singlets decaying to jets, the details of the fragmentation would look very different from a lab-frame fragmentation of the QCD background. This information could be exploited as an additional handle in such searches for Beyond the Standard Model (BSM) physics. Building a tagger from this effect can allow such searches to select signal jets in a momentum range of 200--250~GeV with an efficiency of about 85\% while rejecting about 40\% of QCD background jets, according to Vincia simulations.

Fragmentation differences have been exploited in the past without explicitly defining this source of the effect. Boosted $V$ searches occasionally use track multiplicity as a discriminator between merged $V$ jets and background QCD jets~\cite{ATLAS:2019nat}, which takes advantage of the difference between the $m_V/2$ momentum scale dictating the $V$-jet fragmentation and the TeV momentum scale of the background. In boosted $V$ decays, well-isolated quark jets of a few hundred GeV can be easily produced. As discussed above, such jets can have very different fragmentation profiles compared to those from background QCD processes. Using this difference can be an important additional handle in further enriching samples in interesting signal processes.

\section{Conclusion}

The assumption used throughout LHC data analysis that a hadronic jet can be described by only its momentum and species is overly simplistic. Since the fragmentation occurs in a particular frame that need not be the lab frame, it's not meaningful to talk about the properties of jets individually as separate physics objects. Especially when studying the detailed substructure of jets, sample dependence effects can be significant, as shown here in simulation.

While well understood by QCD theory specialists, this effect removes the foundation of jet individualism used by many experimental tools in collider physics. The training of jet-by-jet taggers should consider the effect of boosted color rest frames, and the language around jet physics should be made more precise. This effect also represents an under-explored opportunity for discriminating jets from boosted color singlet decays, especially in BSM searches.

\section*{Acknowledgements}

We thank Tova Holmes, Karri Folan DiPetrillo, Jessie Shelton, Brian Shuve, Jesse Thaler, Zach Marshall, Nishita Desai, Kate Pachal, Max Swiatlowski, Rikab Gambhir, and Ben Thornberry for incredibly useful discussions and suggestions.
We also thank Patrick Kirchgaeßer, Giordon Stark and the mario-mapyde project \cite{Stark:2023ont}, and Matthew Feickert and the scailfin project for very useful Docker images used throughout this study.
This work has been supported by the Department of Energy, Office of Science, under Grant No. DE-SC0023321 and DE-SC0020267 and the National Science Foundation, under Award No. 2235028.

\bibliographystyle{unsrt} 
\bibliography{jetfragbib}
\end{document}